\documentstyle[amssymb,12pt]{article}
\font\tmsb=msbm10 at12pt
\font\smsb=msbm7
\font\ssmsb=msbm5
\newfam\msbfam
\textfont\msbfam=\tmsb
\scriptfont\msbfam=\smsb
\scriptscriptfont\msbfam=\ssmsb
\def \mth {\fam\msbfam}
\def \Mth#1 {{\mth #1}}

\def \ZZ {{\Bbb {Z}}}

\newcommand \bd {\begin {displaymath}}
\newcommand \ed {\end {displaymath}}

\newcommand \pa {\partial}

\newcommand \la {\lambda}

\newcommand \cL {{\cal L}}
\newtheorem {prop} {Proposition} [section]
\newtheorem {defi} {D\'efinition} [section]

\newtheorem {th} {Th\'eor\`eme} [section]

\title {The Lax operators $\cal L$ of the Benney type equations
 bound with the circle}
\author{A.Balan\\
\'Ecole Polytechnique\\
Centre de Mathematiques\\
 UMR 7640 of CNRS\\
F-91128 PALAISEAU Cedex\\
{\it email}: balan@math.polytechnique.fr}
\date{}

\begin{document}
\maketitle

\abstract{The Lax operators of the Benney type equations are studied
 on the circle. The vectors fields of the Lax operators are showed
 to commute with each other}

\section{The injectivity for the Lax operators $\cL$}

The operator $\cL$ of \cite{EOR} can have the decomposition:
\bd
{\cL} = L + \sum_{\la} {\cL}_{\la},
\ed
\bd
{\cL}_{\la} = \sum_i \phi_{\la,i} (\pa + \la)^{-1} \psi_{\la,i},
\ed
with $\phi_{\la,i}, \psi_{\la,i} \in \cup_n Ker(T- \la I)^n$,
 $T$ is the translation, $I$ is the identity.
\bd
\pa^{-2} = \pa^{-1}\pa^{-1}= \pa^{-1}[\pa,t]\pa^{-1}= t\pa^{-1}-\pa^{-1}t.
\ed
A first question is the one of injectivity.

{\th . 
\bd
\sum_{\la} {\cL}_{\la} = 0 \Rightarrow {\cL}_{\la}=0.
\ed}

It is possible to make the choice of analytic functions around the
unity circle. The injectivity can then be proved.

\bd
\sum_{\la} {\cL}_{\la}=0,
\ed
\bd
{\cL}_{\la} = \sum_i \phi_{\la,i} {\pa}^{-1} \psi_{\la,i},
\ed
chosing a free family in $\cup_n ker(T-\la I)^n$, it give:
\bd
\sum_{\la,i} \tilde \phi_{\la,i} \tilde \psi_{\la,i}^{(n)} \pa^{-n-1}
 a_{\la,i}=0,
\ed
so for all $n$,
\bd
\sum_{\la,i} \tilde \phi_{\la,i}^{(n)} \tilde \psi_{\la,i} =0.
\ed 
Using the Wronskian, it is then possible to show injectivity.

\section{The product of two operators}

It is possible to write generically:

\bd
{\cL}= A^{-1}B= B'A'^{-1}.
\ed

It must be showed:
\bd
\psi_{\la} \tilde \phi_{\mu} \in \cup_u Ker(T-\la^{-1} \mu I)^n.
\ed
The product ${\cL}^2$ can have the same form.
\bd
a(\pa+\la)^{-1}b . c (\pa+ \mu)^{-1}d =
\ed
\bd
= a (\pa+\la)^{-1}[ \pa, \int^x bc](\pa+ \mu)^{-1}d=
\ed
\bd
= a (\int^x bc)( { 1 \over {\pa + \mu}}) d - a (\pa + \la)^{-1}(\int^x bc)
{ {\pa+ \la} \over {\pa + \mu}} d.
\ed
\bd
a(t+1) = e^{\la} a(t),
\ed
\bd
b(t+1) = e^{-\la} b(t),
\ed
\bd
c(t+1)= e^{\mu}c(t),
\ed
\bd
d(t+1)= e^{-\mu}d(t).
\ed
\bd
a \pa^{-1}[ \pa, (\int^x bc)_{\mu - \la}] \pa^{-1}d=
\ed
\bd
= a_{\la} \pa^{-1}(\int^x bc)_{\mu-\la} \pa^{-1}d_{-\mu}-
  a_{\la} \pa^{-1} (\int^x bc)_{\mu - \la} d_{-\mu}.
\ed

{\defi . An operator is of type $\la$ if:
\bd
\sum_i \phi_i \pa^{-1} \psi_i,
\ed
with $\phi_i \in \cup_n Ker(T- \la I)^n, \psi_i \in \cup_n Ker(T-\la^{-1}I)^n$.
}

With this definition, $A$ is of type  $\la$ and $B$ of type $\mu$
 implies that $AB=C+D$ with $C$ of type $\la$, $D$ of type $\mu$.  

{\prop . Let ${\cL}$ be a Lax operator $\ZZ$ periodic, then:
\bd
{\cL} = L + \sum_i \phi_{\la,i} \pa^{-1} \psi_{\la,i},
\ed
with:
\bd
\phi_{\la,i} \in \cup_n Ker(T- \la I)^n,
\ed
\bd
\psi_{\la,i} \in \cup_n Ker(T-\la^{-1}I)^n.
\ed
And the product of two such operators ${\cL}{\cal M}$ 
 has the same form.}

\section{The vectors fields of the Lax operators $\cL$}

{\defi . Vectors fields of the Lax operators ${\cL}$ are defined as:
\bd
\pa_{\la,i} {\cL} = [ {\cL}^i_{\la}, {\cL}].
\ed
}

\subsection{The phase space}

Let ${\cL}_{\la}$ be the Lax operator corresponding with $\la$.
\bd
{\cL}_{\la} = \sum_i \phi_{\la,i} \pa^{-1} \psi_{\la,i}.
\ed
with:
\bd
deg(\phi_{\la,i}) + deg( \psi_{\la,i} \leq N_{\la}.
\ed
\bd
[{\cL}_{\la}, {\cL}_{\mu}] = [ \phi_{\la} \pa^{-1} \psi_{\la},
\phi_{\mu} \pa^{-1} \psi_{\mu}].
\ed
\bd
\pa \phi_{\mu}= \phi_{\la} ( \int^x \psi_{\la} \phi_{\mu}).
\ed
of degree $deg(\phi_{\mu}) + N_{\la} +1$.

\subsection{The commutativity}

{\th . There is commutativity of the vectors fields:
\bd
[ \pa_{\la} - {\cL}_{\la} , \pa_{\mu} -{\cL}_{\mu}]=0.
\ed}

{\it Proof}:

it must be showed:
\bd
\pa_{\la}( {\cL}_{\mu} )- \pa_{\mu} ( {\cL}_{\la}) = [
 {\cL}_{\la} ,{\cL}_{\mu}].
\ed
\bd
\pa_{\la} {\cL} = [ {\cL}_{\la}, {\cL}] 
=\sum_{\mu \neq \la} [ {\cL}_{\la}, {\cL}_{\mu}].
\ed
\bd
\pa_{\la} ( {\cL})_{\mu} =( \pa_{\la} {\cal L})_{\mu}=
 ( [ {\cL}_{\la}, {\cL}_{\mu}])_{\mu}, 
\ed
\bd
\pa_{\mu} ( {\cL})_{\la} =( \pa_{\mu} {\cal L})_{\la}=
 ( [ {\cL}_{mu}, {\cL}_{\la}])_{\la}, 
\ed
\bd
\pa_{\la}( \cL_{\mu}) - \pa_{\mu}( \cL_{\la}) =
 ( [ \cL_{\la}, \cL_{\mu}])_{\mu} + ( [ \cL_{\la}, \cL_{\mu}])_{\la}.
\ed
\bd
(\pa_{\la,i} {\cL}^j)_{\mu} = [ (\cL^i)_{\la}, {\cL}^j]_{\mu}=
\ed
\bd
= \sum_k \cL^k ( \pa_{\la,i} \cL) \cL^{j-1-k}.
\ed
If $\la \neq \mu$,
\bd
[( {\cL}^i)_{\la} , ({\cL}^j)_{\mu}]_{\mu} -
  [( {\cL}^i)_{\mu} , ({\cL}^j)_{\la}]_{\la} = [ (\cL^i)_{\la}, (\cL^j)_{\mu}].
\ed
If $\la= \mu$,
\bd
( [{\cL}^i_{\la} , {\cL}^j] + [( {\cL}^i_{\mu} , {\cL}^j_{\la}])_{\la} =
\ed
\bd
= ( [ - \sum_{\nu \neq \la}( \cL^i)_{\nu}, \cL^j])_{\la}-
\ed
\bd
-([ \sum_{\nu \neq \la}( \cL^i)_{\mu}, (\cL^j)_{\la}])_{\la}+
( [ \cL^i, (\cL^j)_{\la}])_{\la}= ([(\cL^i)_{\la},(\cL^j)_{\la}])_{\la}.
\ed
\subsection{The Hamiltonians}
\bd
{\cL}_{\la}= \sum_{i} \tilde \phi_{\la,i} \pa^{-1} \tilde \psi_{\la,i},
\ed
with:
\bd
\tilde \phi_{\la,i} \in Ker(T- \la I)^n,
\ed
\bd
\tilde \psi_{\la,i} \in Ker(T- \la^{-1}I)^n.
\ed
\bd
res( {\cal L}_{\la}) = \int_{S^1} \sum_i \tilde \phi_{\la,i}
 \tilde \psi_{\la,i}.
\ed
With the choice of $\cL$ $\ZZ$ periodic, the ${\cL}_{\la}$ are
 $\ZZ$ periodic.
\bd
A \cL= B, A^T \cL= B^T,
\ed
gives:
\bd
A=A^T, B=B^T.
\ed
If $L=0$.
\bd
\cL = A_{\la}^{-1}B_{\la} + {\buildrel \approx \over{{A}}}^{-1} \tilde B,
\ed
equivalent with:
\bd
B= \tilde A B_{\la} + C \tilde B.
\ed
\bd
A_{\la} {\buildrel \approx \over{{A}}}^{-1} = \tilde A^{-1}C.
\ed
 \bd
\tilde A  A_{\la}= C \buildrel \approx \over{{A}}.
\ed 
\bd
B= \tilde A B_{\la}+ C \tilde A.
\ed
$A= C {\buildrel \approx \over{{A}}}$
 has for zeros $\phi_{\la,i}$, and $A^*= {\buildrel \approx \over{{A}}}^*
 C^*$ has for zeros $\phi_{\la^{-1},i}^*$.


\begin{thebibliography}{3}
\bibitem{EOR}
B.Enriquez, A.Yu.Orlov et V.N.Rubtsov, {\it Dispersionful analogues of
 Benney's hierarchies}, Inverse Problems 12 (1996),
 p.241-50~; solv-int/9510002.
\end{thebibliography}
\end{document}